**Tracking thermal transport in colloidal quantum dot films using *in-situ* time-resolved X-ray diffraction**

Eliza Wieman[1], Nejc Nagelj[2], Ethan Curling[3], Larry Chen[4], Jin Yu[5], A. Paul Alivisatos[3,6], Aaron Lindenberg[7], Benjamin T. Diroll[8], Jacob H. Olshansky[2], Jihong Ma[9]*, Burak Guzelturk[5]*, Benjamin L. Cotts[1]*

1. Department of Chemistry, Middlebury College, Middlebury, VT, USA
2. Department of Chemistry, Amherst College, Amherst, MA, USA
3. Department of Chemistry, University of California at Berkeley, Berkeley, CA, USA
4. Department of Materials Science and Engineering, Northwestern University
5. X-ray Science Division, Argonne National Laboratory, Lemont, IL, USA
6. Department of Chemistry, University of Chicago, Chicago, IL, USA
7. Department of Materials Science and Engineering, Stanford University, Stanford, CA, USA
8. Center for Nanoscale Materials, Argonne National Laboratory, Lemont, IL, USA
9. Department of Mechanical Engineering, University of Vermont, Burlington, VT, USA

**correspondence to:* jihong.ma@uvm.edu, burakg@anl.gov, bcotts@middlebury.edu

**Abstract**

Colloidal quantum dots (QDs) and their thin-films are increasingly used in electronic and photonic devices replacing traditional bulk semiconductors. However, thermal properties of the QDs are comparatively underexplored relative to device development efforts. This study shows the use of time-resolved X-ray diffraction as a contact-free method to probe the thermal response of QDs in device-like environments, providing *in-situ* insights for future thermal management strategies. Through the extraction of Debye-Waller Factors on a sub-nanosecond timescale, we use time-resolved X-ray diffraction to directly capture the heating and cooling of core/shell CdSe/CdS QDs following pulsed optical excitation. In a QD thin-film that actively provides optical gain, the thermal conductivity is found to be as low as 0.55 W $m^{-1}$ $K^{-1}$, because of the poor heat flow within close-packed QD solids. For QDs dispersed in liquids, interfacial thermal conductance is found to dominate the thermal relaxation with a conductance on the order of 15 MW $m^{-2}$ $K^{-1}$.

## Introduction

As electronic and photonic applications increasingly utilize nanomaterials, including quantum-confined nanocrystals, understanding nanoscale thermal properties and their divergence from bulk properties is critical for device development and performance optimization. The thermal conductivity ($\kappa$) of semiconductor quantum dot (QD) films ranges between 0.1 and 0.8 W $m^{-1}$ $K^{-1}$,[1–6] which is more than an order of magnitude lower than their analogous bulk materials.[1,3,7] As a result, heat buildup in QD layers from nonradiative pathways can erode the performance and lifetimes of optoelectronic devices such as lasers, light-emitting diodes, and photovoltaic solar cells.[1,2,8–10] Similarly, poor thermal dissipation has been considered to impede the realization of QD-based continuous-wave and electrically-injected lasers.[11–14] Conversely, thermoelectric devices, which can take advantage of poor thermal transport in QD solids, are under development to convert thermal gradients into electricity.[1,8,9]

While the relationship between thermal transport and device performance is studied extensively in bulk or epitaxially-grown semiconductor systems[15], the study of thermal transport in QD assemblies lags behind their device development. Particularly, accessing temperature changes on ultrafast timescales (pico- to nanosecond) concomitant to excited electronic/excitonic processes remains a challenge. Conventional methods for measuring thermal conductivity include 3ω[16] and time or frequency domain thermoreflectance[2,17], which have been used to determine thermal transport in a broad range of materials, including QD solids.[2–5] However, these techniques commonly require deposition of a metal transducer layer on the material of interest to probe temperature changes indirectly. The introduction of a transducer also complicates the quantification of the intrinsic thermal response of the underlying sample due to the formation of the sample/transducer interface with attendant interfacial thermal conductance.

By carefully monitoring changes in Bragg peak positions and intensities, time-resolved X-ray diffraction (TR-XRD) directly measures the changes in bond length and mean-squared atomic displacements linked directly to the transient heating and cooling of materials.[18–20] TR-XRD has been employed to investigate thermal transport in various nanoscale materials, such as layered two-dimensional systems.[18–20,24–26] Nevertheless, application of this technique to QDs has remained limited. Several TR-XRD studies have previously reported non-thermal structural responses in various nanocrystal systems, such as surface melting in CdSe QDs, structural phase transitions in $CsPbBr_3$ QDs, and local symmetry changes in PbS QDs.[21–23,27–34] These TR-XRD experiments typically study nanocrystals in dilute liquid environments; however, such conditions are incompatible with realistic QD device architectures, which employ solid thin-films. To this end, there is a need to adapt the TR-XRD approach to quantify thermal transport in device-like environments of the QDs.

In this work, we use TR-XRD to study the thermal transport of prototypical core/shell CdSe/CdS QDs in both thin-films and liquid forms. We utilize the reflection-type grazing-incidence wide-angle X-ray scattering (GIWAXS) geometry to maximize the diffraction signal from QD thin-films deposited on a substrate. In the case of samples dispersed in a liquid jet, we use a transmission mode wide-angle X-ray scattering (WAXS) geometry. Using both geometries (see Figure 1), we access the thermal response of QDs following above-bandgap pulsed laser excitation causing impulsive heat generation and subsequent thermal relaxation. We validate the thermal response by measuring transient Debye Waller Factors, which cause systematic decrease of Bragg peak intensities as a function of scattering vector magnitude, and we perform finite element thermal modeling to quantitatively extract transport parameters.

In a QD thin-film exhibiting active optical gain, mimicking the excitation regime required for a QD laser, we find the thermal relaxation to be many orders of magnitude slower than those of individual QDs in a dilute liquid dispersion. This substantial difference in the thermal dissipation timescale ($10^{-6}$ s in thin-film vs. $10^{-10}$ s in liquid) elucidates the severity of the slow thermal dissipation across close-packed QD solids conventionally used in photonic devices. We quantify the thermal conductivity of such laser-like QD thin-films to be 0.55 W $m^{-1}$ $K^{-1}$ by modeling the TR-XRD data. In the case of liquid samples, thermal relaxation is primarily governed by QD-ligand/liquid interface with an interfacial thermal conductance on the order of 15 MW $m^{-2}$ $K^{-1}$. Overall, TR-XRD stands out as a quantitative method to investigate thermal transport in device-like environments of semiconductor QDs, which can be also extended to other nanomaterial solids. With further refinement, this methodology can be extended to *operando* conditions, which could then help tackle thermal management issues in photonic and quantum devices of nanomaterials.

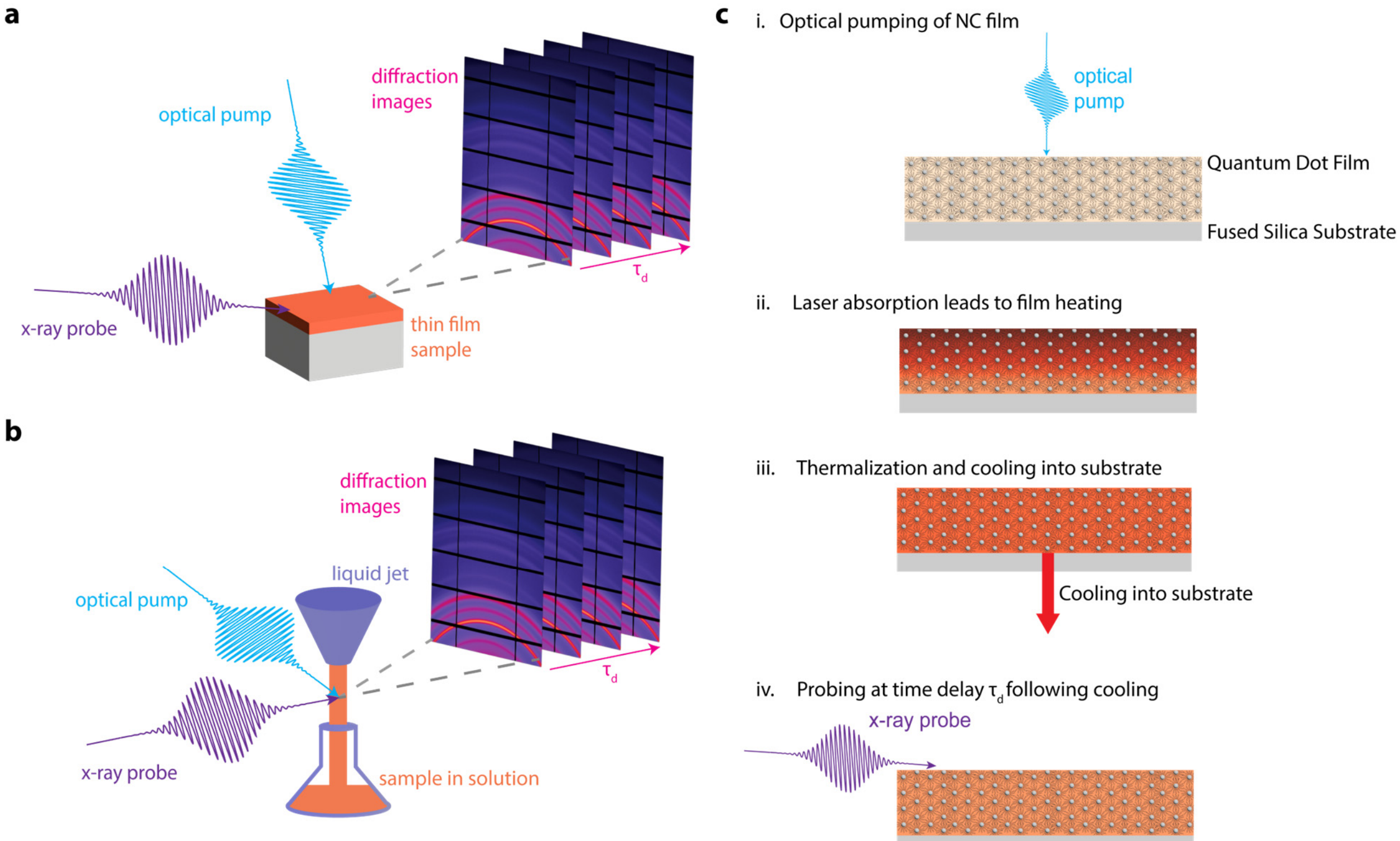

**Figure 1. Time-resolved x-ray diffraction of CdSe/CdS QDs.** Schematic of the time-resolved X-ray diffraction experiment, where a sample is excited with a 400 nm optical pump and then probed with X-rays at multiple pump-probe delays to measure photoinduced structural changes in **(a)** thin-film or **(b)** liquid jet samples. **(c)** Schematic of the thin-film thermal dynamics. The laser pulse is absorbed at $\tau_d = 0$ heating the film through hot-carrier cooling and fast nonradiative relaxation. Following the initial thermalization, the film is probed at pump–probe delay $\tau_d$ to determine the nanocrystal temperature, averaged across the film thickness and beam footprint, thereby tracking the film cooling timescales.

## Results and Discussion

Due to high photoluminescence quantum yields, improved environmental stability, and well-established synthesis, we study core/shell CdSe/CdS QDs as a model system here. We synthesize these QDs based on previous reports (See SI section A for synthesis methods) [35,36]. The sample of interest is a core/shell QD with a core size of 3.5 nm and a total mean diameter of 9.4 nm. Another batch of samples with a mean diameter of 11.6 nm has also been tested and showed consistent results. All the QD samples used here have oleic acid ligands on their surfaces.

For thin-film samples, we either spin coat or drop cast the core/shell QDs on a fused silica substrate and let the solvent evaporate. This process leads to randomly-oriented and close-packed QD thin-films with a thickness in the range of 100 – 1000 nm. In the case of liquid samples, we transfer these QDs from toluene to dodecane solvent, as dodecane has a higher boiling point which improves the robustness of the liquid-jet based TR-XRD measurements.

We use 3.1 eV photon energy pulsed optical excitation in both the liquid jet and thin-film TR-XRD experiments. This excitation is above the bandgap of the QDs, hence causing an impulsive heating through fast hot-carrier cooling as well as Auger nonradiative recombination channels. After photoexcitation, the QDs reach a thermal equilibrium state within tens of picoseconds[36]. We then use TR-XRD to monitor the thermal relaxation kinetics of the QDs over nanosecond to microsecond timescales.

**Thermal transport in device-like environment**

To establish if the thin-film sample can be considered as a *de facto* optoelectronic device, we probe the optical emission under varying excitation fluence in the same geometry used for the GIWAXS experiment (Figure 1a; See SI section F for measurement methods). Under low excitation fluence, the emission spectrum shows a broad peak at around 665 nm due to spontaneous emission of the single exciton state. Upon increasing excitation fluence, we observe the thresholded emergence of a new emission peak at around 625 nm (Figure 2a). This blue-shifted sharper peak arises due to amplified spontaneous emission (ASE), which results from a net optical gain within the thin-film. The net optical gain arises from the stimulated emission among repulsive biexciton (XX) states in this quasi Type-II CdSe/CdSe core/shell system[37]. The ASE threshold (i.e. inflection point in Figure 2b) in this 600 nm thick thin-film is 55 μJ $cm^{-2}$, which is consistent with thresholds reported in similar CdSe/CdS core/shell QDs[38,39].

ASE is the precursor of a laser as the thin-film acts as a weakly-coupled waveguide with broad cavity modes. Under increased excitation fluence, the sample even exhibits a higher-order optical gain from p-state excitons, resulting in a new ASE peak at shorter wavelength (575 nm)[40]. The slope of the ASE peak as a function of excitation fluence represents the differential optical gain and the ASE slope is much higher than that of the PL peak due to the contribution of stimulated emission. Overall, our sample matches the form factor of a QD laser active layer and captures its key optical-gain behavior.

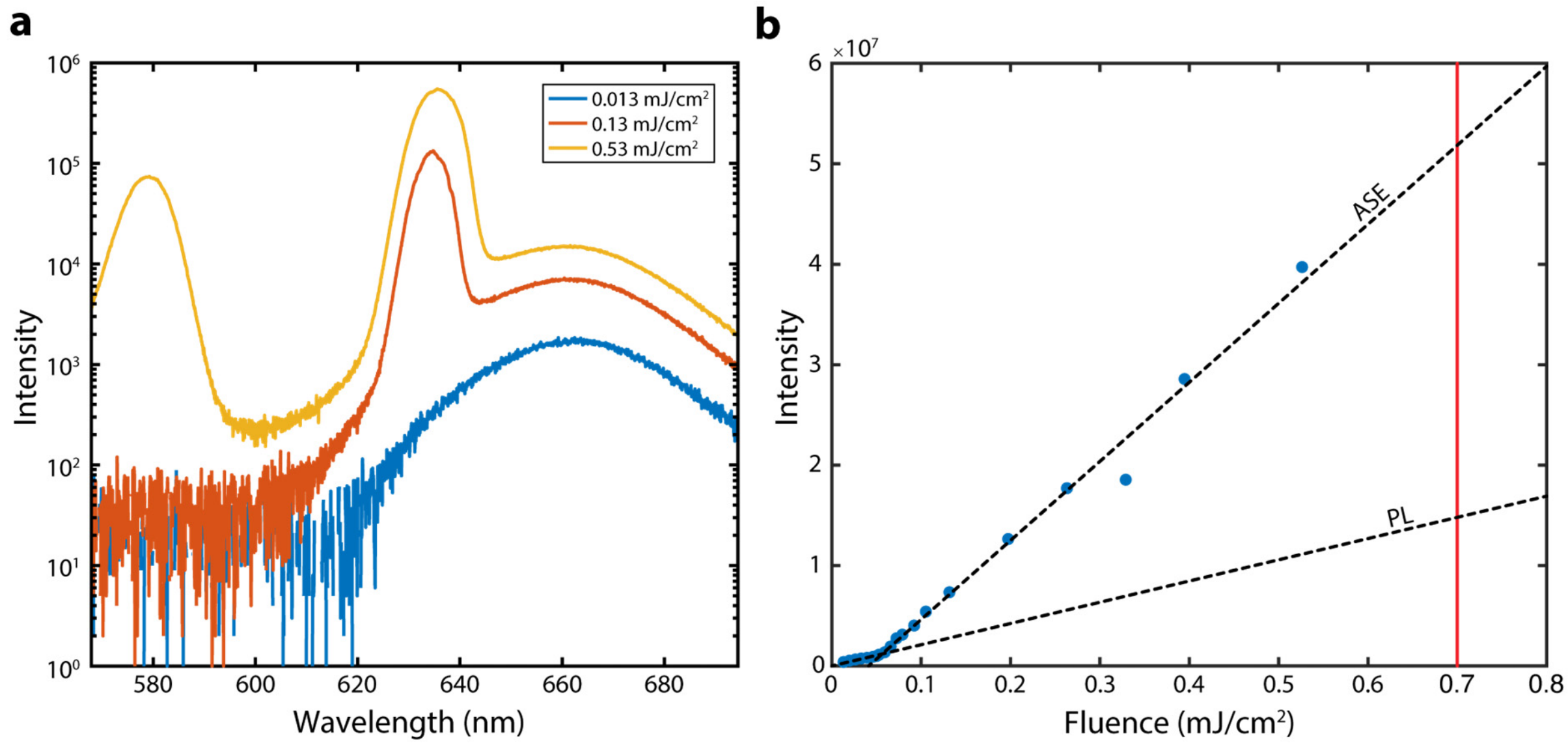

**Figure 2. ASE in thin-film sample. (a)** Log-linear plot of spectra from fluence-dependent optical emission experiments, showing emission above (yellow, orange) and below (blue) ASE threshold fluence. At higher fluences, an additional ASE mode is observed, indicated by an additional peak at ~575 nm. **(b)** Integrated emission intensity versus fluence for all recorded fluences. Data was fit to determine an ASE threshold of approximately 0.056 mJ/cm$^2$ at the selected spot in the film, well below the fluence used in TRXRD experiments (0.7 mJ/cm$^2$), shown as a red vertical line.

Having established the ability of our films to act as a model optoelectronic device, we perform time-resolved GIWAXS measurements under ASE conditions to assess the dynamic thermal response during optical gain (see SI section C for methods). Figure 3a shows the two-dimensional diffraction image exhibiting Debye-Scherrer rings, which we radially integrate to obtain one-dimensional diffraction intensity I(Q), where Q is the scattering vector magnitude. Figure 3b shows the static diffraction pattern of a thin-film sample prior to optical excitation, $I_0(Q)$, and the change in diffraction intensity 50 ns after the pump pulse, $\Delta I(Q, t = 50\text{ ns})$. The $\Delta I(Q)$ shows a derivative-like pattern at all major Bragg peaks, indicating a shift of Bragg peaks to lower Q values, which arises from lattice expansion associated with heating. The derivative signals also skew toward negative lobes, especially at increased Q. This indicates that the total Bragg peak intensities also decrease particularly at higher Q. Such a decrease in Bragg peak intensity is consistent with the Debye-Waller factor arising from transient heating. A similar pattern is observed in liquid jet data as seen in Figure S2.

To decode the thermal response, Debye-Waller factors (DWF) were extracted as a function of temporal delay between laser pump and X-ray probe pulses. DWF quantifies the Q-dependent changes in diffraction peak intensities associated with the mean-squared atomic displacements associated with thermally-induced lattice motion, or random disorder. In the most basic representation, DWF = $e^{-\langle[\boldsymbol{Q} \cdot \boldsymbol{u}]^2\rangle}$ where $\boldsymbol{Q}$ and $\boldsymbol{u}$ are scattering and atomic displacement vectors, respectively. DWF relates the Bragg peak intensities at different Q to the underlying atomic displacements.[23] Under the assumption of isotropic, harmonic displacements of atoms around their equilibrium positions due to thermal energy, the DWF leads to a logarithmic relationship between $I(Q)$ and $Q^2$, expressed as $-\ln(I/I_0) = \frac{Q^2}{3}\langle u^2\rangle$. Here, $I_0$ is diffraction intensity at 0 K, $I$ is the diffraction intensity at a given temperature, $Q^2$ is the scattering vector squared, and $\langle u^2\rangle$ is the mean-squared atomic displacements. The DWF relationship can be expressed for a pulsed laser excitation experiment for a pump-probe delay, $t$, $-\ln(\mathrm{I(t)}/I_0) = \frac{Q^2}{3}\langle\Delta u(t)^2\rangle$, where $\langle\Delta u(\mathrm{t})^2\rangle$ is the induced atomic mean-squared displacements due to a change in temperature $\Delta T(t)$ at the time delay $t$.[36] This relationship is plotted as $-\ln(I(t)/I_0)$ vs $Q^2$ for both liquid and solid-state data, with the linear trend observed in both samples indicative of transient heating with the slope being equal to $\langle\Delta u(t)^2\rangle$ (Figure 3c, d). The linear best fit slope is $5.1 \times 10^{-4}$Å$^{-2}$ for the liquid jet data at 50 ps (Figure 3c) and $4.1 \times 10^{-4}$ Å$^{-2}$ for thin-film data at 50 ns (Figure 3d), which corresponds to temperature increases of 6.8 K and 5.5 K respectively (SI Section C).

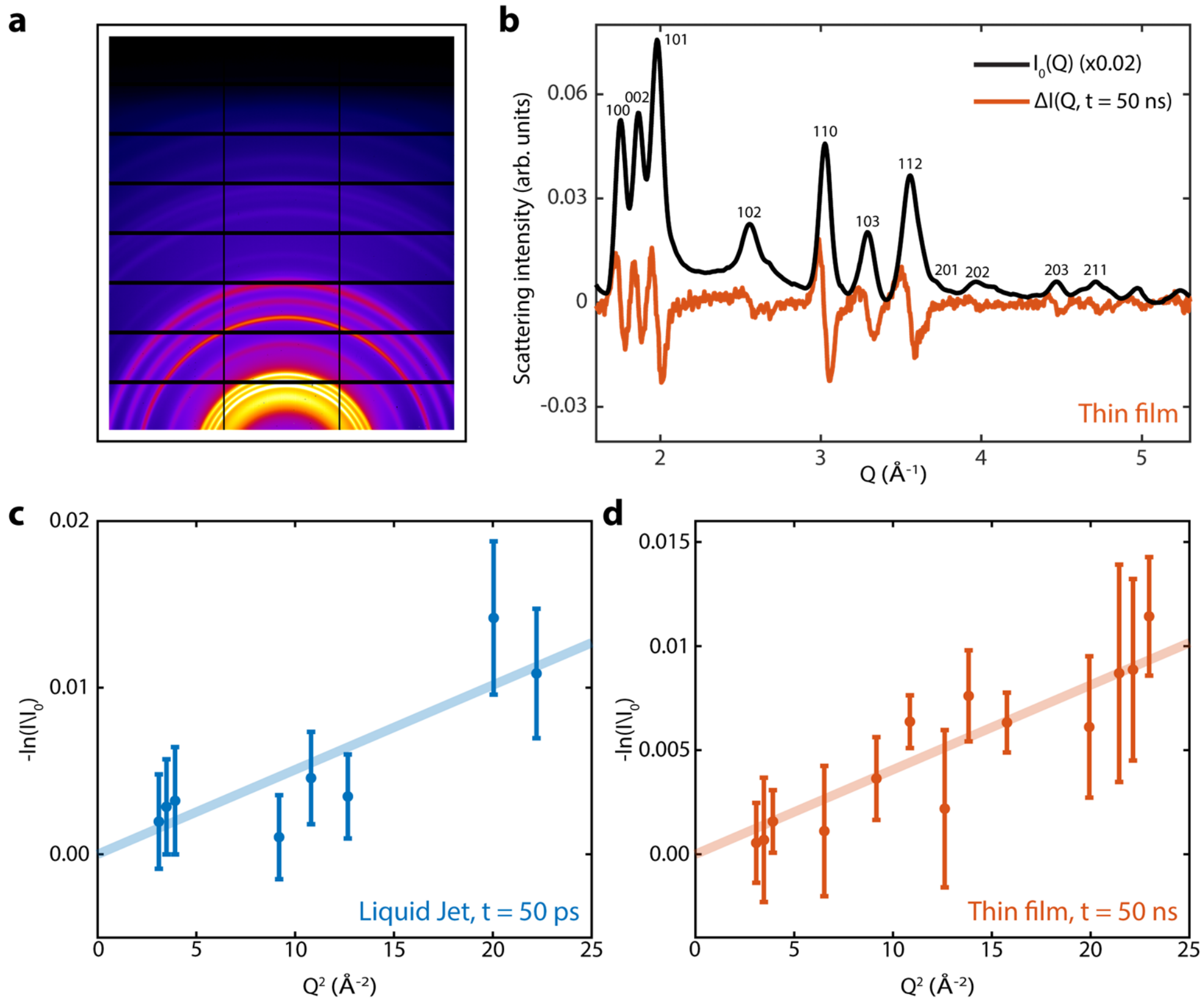

**Figure 3. Debye–Waller analysis of transient diffraction intensity changes. (a)** Representative diffraction image of the CdSe/CdS thin-film sample. **(b)** Radially integrated, background-subtracted diffraction intensity of a CdSe/CdS thin-film sample prior to excitation, $I_0(Q)$, scaled to 2% (black), and the transient change in diffraction intensity 50 ns after excitation, $\Delta I(Q, t = 50\text{ ns})$ (orange). **(c, d)** Debye-Waller Factor (DWF) plots of liquid jet and solid state samples at t = 50 ps and t = 50 ns, respectively. Error bars represent standard error of the mean. The linear relationships observed in both plots are indicative of transient lattice heating (DWF). The higher signal-to-noise ratio in the solid-state data enables more diffraction peaks to be resolved in the TR-XRD spectra, resulting in a more robust DWF analysis.

Thin-film samples make the extraction of transient DWF easier compared to liquid jet form for several reasons. First, thin-films enable better penetration depth matching in the surface-sensitive GIWAXS geometry because both the optical pump and X-ray probe penetrations are effectively limited to the sample thickness. In addition, the excitation density is also more homogenous across the sample thickness, as film thickness is comparable to the optical penetration depth on the order of 100–1000 nm.

Extracting DWF from liquid jet samples presents some challenges. Liquid jets are typically much thicker (500–1000 μm), making pump-probe penetration depth matching harder to satisfy and producing a stronger gradient in QD excitation density across the jet. Additionally, the commonly adopted liquid jet configuration generally requires high QD concentrations to generate strong X-ray diffraction intensity. Despite increasing the static diffraction signal, more concentrated samples absorb the incoming laser power in a thinner region, reducing the number of relative excitations over the jet depth and lowering the overall differential signal (Figure S4). Jet and sample density fluctuations further contribute large background noise to $I(Q)$, complicating robust extraction of the DWF ($-\ln(I(t)/I_0)$). Together these factors likely explain the lack of reports of DWF in liquid NC samples analyzed by prior TR-XRD measurements.[21–23,27,30,41] Consistent with the lower signal to noise, Bragg peaks with differential signal comparable to background fluctuations have been eliminated from the liquid jet data in Figure 3c. The higher signal to noise in the thin-film data allows weaker Bragg peaks to be resolved, yielding more data points in Figure 3d.

The average increase in temperature (ΔT) per QD is set by the fraction of photoexcited electronic energy that is transferred to the atomic lattice via nonradiative relaxation, which scales with the average number of excitations (⟨N⟩) created per QD. The magnitude of the DWF effect is directly related to the lattice temperature of the QDs, allowing tracking of temperature change in QDs following photoexcitation over time. At early time delays, maximum ΔT values of 5.5 K and 6.8 K were extracted from solid and liquid DWF plots, respectively, using a previously determined calibration between $\langle\Delta(u(t)^2\rangle$ and ΔT (See Figure 3c/d and SI Section C).[36] The observed ΔT values are similar despite the different fluences used to excite liquid jet (5 mJ/cm$^2$) and thin-film (0.7 mJ/cm$^2$) samples. We calculate ⟨N⟩ through the jet or film thickness based on incident laser fluence and QD absorption cross-section. Then, we estimate ΔT using the volumetric heat capacity and assuming full conversion of all excitations larger than single excitation state into heat through Auger-Meitner recombination (See SI Section E). As the laser is attenuated with depth, both ⟨N⟩ and the resulting ΔT decreases deeper within the film (See Figures S3-S4). Depth-averaged ΔT values are calculated to be 13 K and 20 K for thin-film and liquid jet samples, respectively. These values overestimate the temperature jumps, as in the actual samples the multi-exciton states have non-zero emissive contribution through spontaneous and stimulated emission channels. As Figure 2 shows, the ASE occurs in the thin-film sample above a threshold fluence of 0.055 mJ/cm$^2$ below the experimental fluence of 0.7 mJ/ cm$^2$ (see also Figures S7).

In liquid jet data the discrepancy in $\Delta T$ is attributed to heterogeneity in the excitation density across the liquid jet (see Figure S4), which causes nonlinear heating effects at shallow jet depths and from unexcited particles at greater jet depths contributing noise and depleting differential intensity signal (see SI Section E; Figure S4). Furthermore, given the available time resolution of the experiments (~100 ps due to the X-ray source), QDs start cooling rapidly in the liquid jet samples, resulting in a lower $\Delta T$ at the first measurable time point.[42]

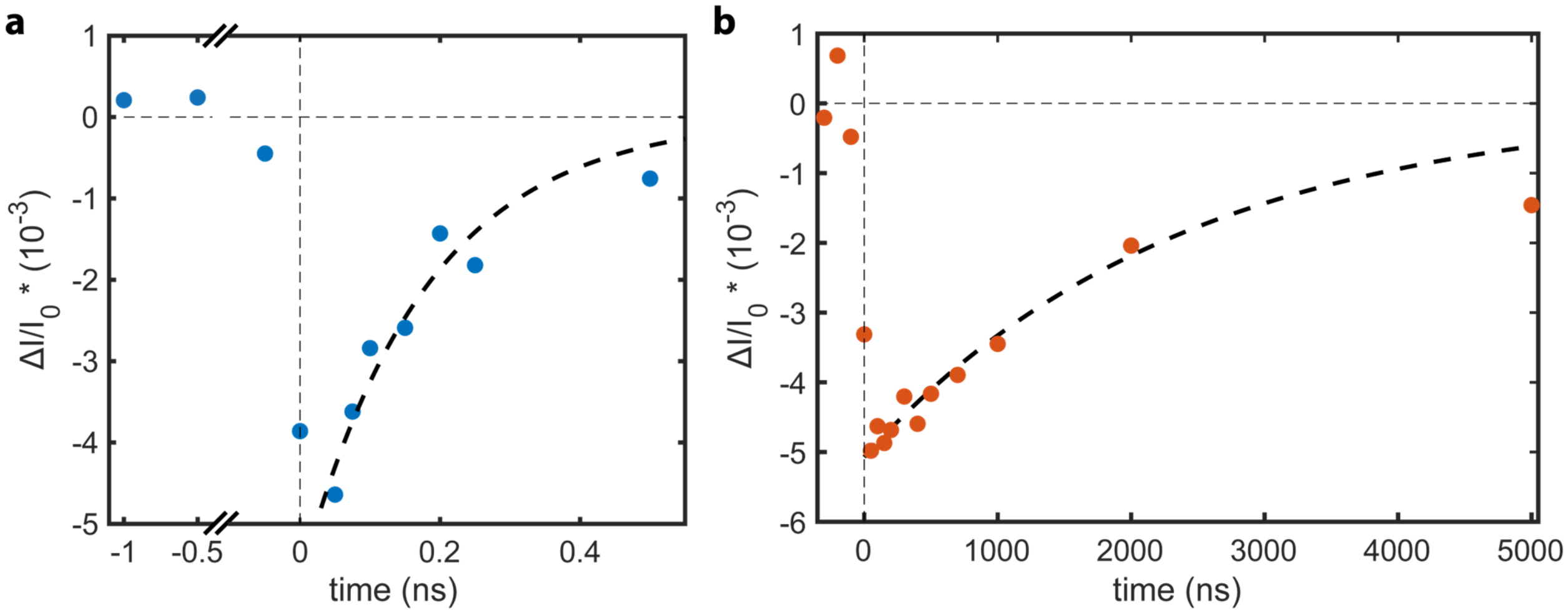

**Figure 4. Temporal recovery of diffraction intensity**. **(a)** Liquid-jet measurements of $\Delta I/I_0(t)$ or the averaged response from the <110> and <103> diffraction peaks of CdSe:CdS nanocrystals suspended in liquid dodecane. An x-axis break compresses the pre-time-zero baseline. A single exponential fit ($ae^{-t/\tau}$) in black dashed line yields a lifetime of $\tau$ = 180 ps. **(b)** Thin-film measurements of of $\Delta I/I_0(t)$ for the averaged response from the <110> and <103> diffraction peaks of CdSe:CdS nanocrystals on fused silica substrate. A single-exponential fit ($ae^{-t/\tau}$) in black dashed line yields a lifetime of $\tau$ = 2.3 μs.

The DWF effect also provides an experimental handle to monitor the relative timescales of cooling between solid-state and liquid samples. Thermal recovery data was fit with a single exponential decay ($ae^{-t/\tau}$), giving respective lifetime values of 2.3 $\pm$ 0.7 μs for the thin-film sample and 180 $\pm$ 80 ps for the liquid sample (Figure 4). Strikingly, the recovery times are four orders of magnitude faster in the liquid jet samples compared to thin-films. This difference arises from the different nature of heat flow in both sample systems. Such large differences in recovery time are known to have dramatic impacts on the amount of heat buildup in QD-based devices, affecting device lifetime and performance.[2,8,43] For QDs dispersed in a liquid solvent, we effectively probe the cooling of isolated individual QDs: heat is first exchanged at the QD/ligand interface and then dissipates into the solvent across the QD–ligand/solvent interface. This type of thermal recovery proceeds through interfacial vibrational coupling at each interface. We estimate the interfacial thermal conductance ($G$) of the QD-ligand/solvent interface through a model developed for spherical metallic particles in a liquid[44,45]. Here $G = \frac{rC_L}{3\,\tau_{relax}}$, where $r$ is the particle radius, $C_L$ is the volumetric heat capacity, $\tau_{relax}$ is the thermal relaxation time constant. Thus, we estimate the $G$ to be 15 MW $m^{-2}$ $K^{-1}$ in the CdSe/CdS QD with oleic acid ligands in dodecane solvent, consistent with past all-optical experiments[46].

While TR-XRD in liquids elucidates the cooling timescale of individual QDs when dispersed in a solvent, the thin-film results are heavily affected by the complex heat transport across the QD solid that comprises many inorganic and organic interfaces of the QD inorganic part and organic ligands, hence a mixed hybrid organic-inorganic effective medium for thermal transport. To extract the thermal conductivity in the thin-film sample, we consider a thermal relaxation model for an effective QD solid medium and quantify its thermal conductivity.

**Extracting thermal conductivity by modeling thermal relaxation**

Thin-film thermal conductivity (κ) was determined by fitting recovery data of the thin-film sample with a finite element model (FEM) based on the one-dimensional transient heat transfer partial differential equation (see SI section G and H). The simulated temperature distribution in the thin film as a function of depth and time is shown in Figure 5a. The normalized temperature, $T_n(x,t)$, is defined as, $T_n(x,t) = \frac{T(x,t)-T_\infty}{T(x=0,t=0)-T_\infty}$, where $T(x,t)$ represents temperature as a function of thin-film depth, $x$, and time, $t$, and $T_\infty$ is the ambient environmental temperature. The initial temperature throughout the thin-film follows an exponential decay from linear absorption through the 600 nm film thickness, calculated from ⟨N⟩ as described above (also see SI Section E; Figure S3). Initially, the top of the film is significantly hotter than the deeper layers, with heat redistributing within the film before diffusing into the substrate on a microsecond timescale. The depth-averaged temperature across the QD thin-film, $T_{n,avg}(t)$, is calculated from the simulated temperature distributions to enable comparison with the experimentally measured temperature evolution over time. In Figure 5b, the first experimentally measured temperature at 50 ns is used as the reference temperature to normalize this depth-averaged temperature $T^*_{n,avg}$, defined as $T^*_{n,avg}(t) = \frac{T_{n,avg}(t)-T_{n,i}}{T_{n,i}}$, where $T_{n,i} = T_{n,avg}(50\ ns)$. The simulated data from the FEM model are normalized in the same manner, and the thermal conductivity is extracted by fitting the simulated decay to the experimental data. This yields a thermal conductivity of 0.55 W $m^{-1}$ $K^{-1}$ for the solid state CdSe/CdS QD thin-film (Figure 5b). The extracted thermal conductivity is consistent with values reported for similar QD thin-films measured with other experimental techniques.[2–5] However, this thermal conductivity in QD form is more than an order of magnitude smaller than those of bulk CdSe (κ =9 W $m^{-1}$ $K^{-1}$)[3] and CdS (κ =16 W $m^{-1}$ $K^{-1}$)[7], highlighting the impact of the QD form factor on the thermal conductivity.

Additional fitting to an analytical model of heat transfer, assuming the substrate as a semi-infinite plate, was performed to further corroborate thermal conductivity values. The analytical heat transfer model treats the sample as a semi-infinite slab with an energy pulse on the surface at *t*=0 and assumes full conversion of laser energy to heat (see SI Section E). Fitting the analytical model yields κ = 0.31 W $m^{-1}$ $K^{-1}$, which is lower than the FEM model and previous reports of similar-sized NCs,[3–5] but still falls within the range of conductivity values reported for similar QDs between 0.1 and 0.8 W $m^{-1}$ $K^{-1}$.[2] While both values are feasible, the more rigorous FEM model returns a value closer to that of similar-sized QDs studied with the 3ω technique. This could be due to the fact that the semi-infinite assumption made for the analytical model is barely valid at $t = 5 \times 10^3$ns when the 90% thermal penetration depth, $\delta_p$, is 250 nm, which is already nearly half of the QD film thickness, 600 nm. As time elapses, $\delta_p$ exceeds the QD thickness, invalidating the semi-infinite plate assumption. Thus, the analytical predictions of κ are only accurate when used to fit the experimental data within a small *t*. Given the large uncertainties of experimental measurement, a longer period of measured data is necessary. Hence, the FEM model is more reliable overall. Overall, this work adds to existing knowledge of nanoscale thermal properties, while also highlighting TR-XRD as a viable new method for determining thermal conductivity of nanomaterials, posing some advantages over more established methods.

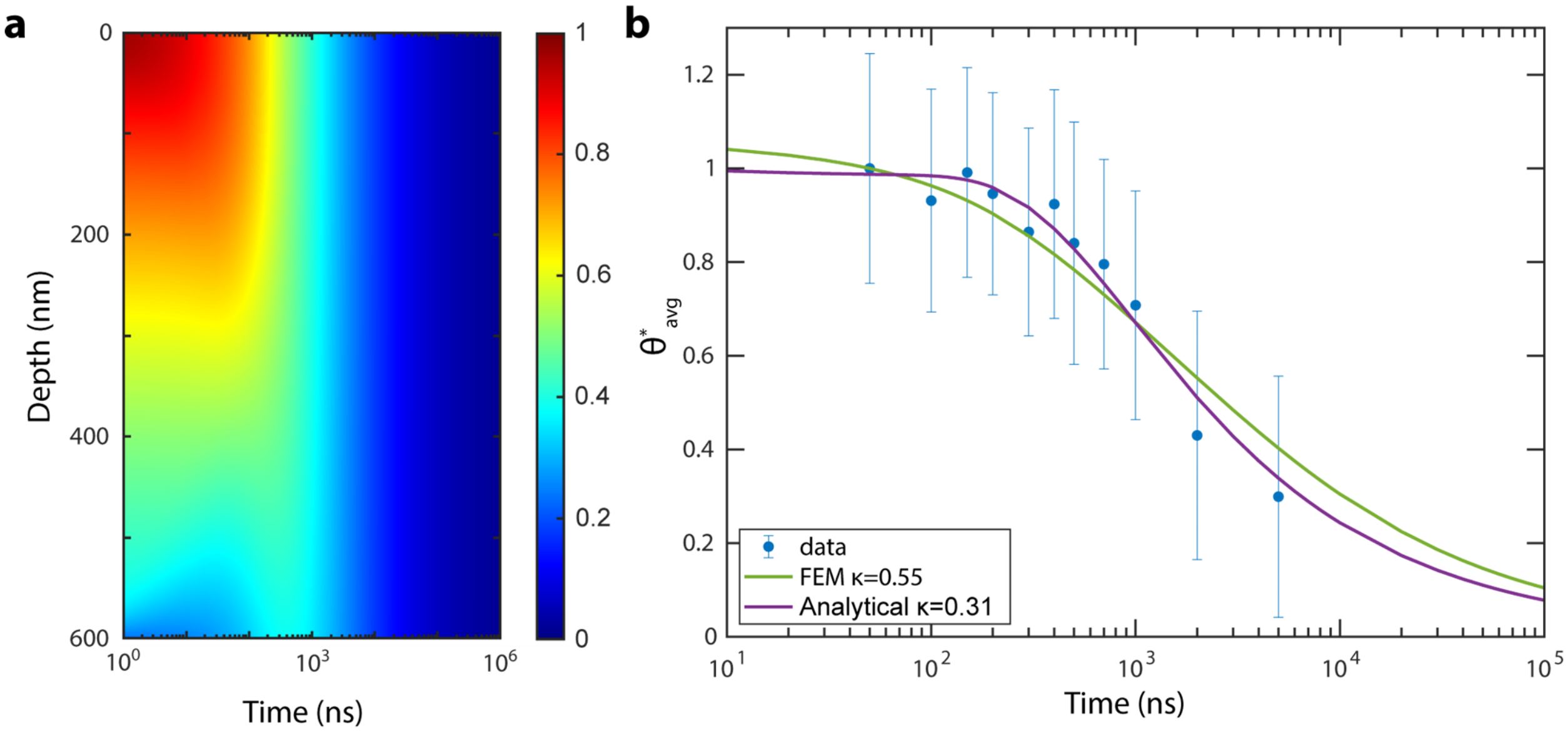

**Figure 5.** Thermal recovery dynamics. **(a)** Simulated film normalized temperature, $\theta$, as a function of film depth and time. The *x*-axis is in log scale. **(b)** Normalized experimentally measured averaged thin-film temperature (blue circles) fit to FEM (green) and analytical (purple) model, yielding fitted thermal conductivity values of 0.55 W $m^{-1}$ $K^{-1}$ and 0.31 W $m^{-1}$ $K^{-1}$, respectively.

Based on the thermal model, we estimate what would be the heating level of our QD thin-film under a 3.1 eV continuous-wave pump if we were to attempt an optically pumped CW laser. We estimate that the sample temperature would rise by about 100 K within 5 µs under realistic laser threshold fluences on the order of 10 kW/$cm^2$ (See SI Section I)[12,14]. This fluence is reasonable considering the 55 $\mu J\ cm^{-2}$ ASE fluence and 20 ns single exciton excited state lifetime. Considering the decomposition, melting and ligand detachment problems at elevated temperatures, active thermal cooling strategies would be imperative to reach stable and robust CW or electrically-injected QD laser operation or other excitation-intensive devices, such as solid-state lighting with QD emitters[14,47].

In conclusion, we have shown that TR-XRD can be used *in-situ* to measure even subtle thermal responses in photoexcited QDs, in both solid and liquid jet forms. While TR-XRD studies are generally performed on liquid samples, this work suggests that solid-state samples produce high quality data that directly informs thermal management in QD-based devices. Additionally, TR-XRD was used to determine thermal conductivity of core/shell CdSe/CdS QD thin-films, adding to a growing body of literature examining nanoscale thermal properties and introducing a new method for measuring thermal conductivity of nanomaterials. Further development and application of TR-XRD based methods for studying nanoscale thermal transport will allow for a more comprehensive understanding of heat flow in QD-based devices, informing device design for a wide range of applications including optoelectronics and thermoelectrics.

**Acknowledgements**
This research used resources of the Advanced Photon Source, a U.S. Department of Energy (DOE) Office of Science user facility operated for the DOE Office of Science by Argonne National Laboratory under Contract No. DE-AC02-06CH11357. Work performed at the Center for Nanoscale Materials, a U.S. Department of Energy Office of Science User Facility, was supported by the U.S. DOE, Office of Basic Energy Sciences, under Contract No. DE-AC02-06CH11357. This material is based upon work supported by the Department of Energy, Office of Science, Office of Basic Energy Sciences, under DE-SC0023425."